\documentclass[prc,twocolumn,floatfix]{revtex4-1}

\usepackage{graphicx}% Include figure files
\usepackage{amssymb,amsmath,amstext,amsthm,amsfonts}
\usepackage{slashed}
\usepackage{epstopdf}

\usepackage[usenames]{color}
\definecolor{magenta}{cmyk}{0,1,0,0}

\definecolor{blue}{rgb}{0,0,1}

\begin{document}

\title{Energy Reconstruction in the Long-Baseline Neutrino Experiment}
\author{U. Mosel}
\email[Contact e-mail: ]{mosel@physik.uni-giessen.de}
\author{O. Lalakulich}
\affiliation{Institut f\"ur Theoretische Physik, Universit\"at Giessen, Giessen, Germany}
\author{K. Gallmeister}
\affiliation{Institut f\"ur Theoretische Physik, Johann Wolfgang Goethe-Universit\"at Frankfurt, Frankfurt a.\ M., Germany}

\begin{abstract}
The Long-Baseline Neutrino Experiment aims at measuring fundamental physical
parameters to high precision and exploring physics beyond the standard model.
Nuclear targets introduce complications towards that aim. We investigate the uncertainties in the energy reconstruction, based on quasielastic
scattering relations, due to nuclear effects. The reconstructed event distributions as a
function of energy tend to be smeared out and shifted by several 100 MeV in
their oscillatory structure if standard event selection is used. We show that a more
restrictive experimental event selection offers the possibility to reach the accuracy needed for a determination of the mass
ordering and the $CP$-violating phase. Quasielastic-based energy reconstruction could thus be
a viable alternative to the calorimetric reconstruction also at higher energies.
\end{abstract}

\date{\today}

\maketitle

\section{Introduction}
The Long-Baseline-Neutrino-Experiment (LBNE) plans to produce a strong neutrino beam from Fermilab near Chicago to a liquid argon detector at the Sanford Underground Research Facility, in Lead, South Dakota, about 800 miles away \cite{Adams:2013qkq}. By comparing the event rates as a function of neutrino energy at the Sanford Underground Research Facility with those at Fermilab the oscillation parameters can be extracted \cite{Adams:2013qkq,Bass:2013taa}. While all mixing angles are now roughly known the experiment aims for a more precise determination of these angles and, in particular, for a determination of the mass hierarchy and the so far undetermined $CP$-violating phase.

The determination of the oscillation parameters depends on the knowledge of the neutrino energy. This energy has to be reconstructed on an event-by-event basis because the neutrino beams are quite broad in energy due to their production in a secondary decay of primarily produced hadrons. The LBNE group has opted for a calorimetric energy reconstruction method; its difficulties lie in experimental limitations such as acceptances and detection efficiencies \cite{Harris:2003si,Kordosky:2007tu,Dytman:2008zz}. Many lower-energy experiments have instead determined the neutrino energy from the kinematics of the outgoing lepton alone by assuming quasifree kinematics for quasielastic scattering (QE) on a neutron at rest. In this method complications arise from the fact that all experiments nowadays use nuclear targets (Ar for LBNE). First, the neutron is not free and not at rest, but instead Fermi moving inside a nuclear potential well. Second, because of final state interactions non-QE events may be misidentified as being QE; for these events the energy is reconstructed from an expression that is not correct. For the simplified situation at lower energies, where only QE and pion production play a role, Fermi motion leads to a distribution of reconstructed energy around the true neutrino energy. Furthermore, so-called "stuck (=reabsorbed)-pion" events produce a lower-energy bump in the reconstructed energy distribution \cite{Leitner:2010kp}. While pion production is the major background to any QE scattering event it has later been shown that also the so-called $2p-2h$ excitations \cite{Martini:2012fa,Nieves:2012yz,Martini:2012uc, Lalakulich:2012ac,Benhar:2013oba} and all other reaction processes \cite{Lalakulich:2012hs} lead to a downward shift of reconstructed energy. The effect of these uncertainties in the reconstructed energy on the extracted oscillation parameters was explored in Refs.\ \cite{Meloni:2012fq,Lalakulich:2012hs,Coloma:2013rqa,Coloma:2013tba} for the MiniBooNE and T2K experiments.

For a higher-energy experiment, such as LBNE, where different reaction processes contribute, no information on the accuracy that can be reached in the QE-based energy reconstruction is available. It is, therefore, the purpose of the present paper to perform such an analysis for the planned LBNE; this experiment will work with a flux with a maximum at about 2.5 GeV with a long high-energy tail, up to 30 GeV. Inspection of the detailed studies of the physics potential of LBNE in \cite{Adams:2013qkq} shows that an energy resolution of about 100 MeV is needed to resolve the region between the first and second oscillation maxima. It is, therefore, the purpose of this Letter to explore how close the QE-based energy reconstruction method can come to this limit and if it could offer a viable alternative to the calorimetric method also at the higher energies of LBNE.

\section{Method}
We use the transport model GiBUU to model both the initial and the final state interactions \cite{Buss:2011mx}. This model has been widely used and tested in very different physics scenarios, ranging from a description of heavy-ion reactions to photon-, electron- and neutrino-induced reactions. It combines the initial reaction mechanisms true QE scattering, $2p-2h$ excitations and pion (and other meson) production through resonances, background processes and deep inelastic scattering (DIS) with a well-tested description of final state interactions. The QE scattering uses an axial mass $M_A = 1$ GeV and employs a mean-field spectral function based on a local, bound Fermi-gas model. Pion production is in its vector interaction part consistent with the MAID analysis for electron scattering and DIS is treated via the high-energy hadronization model PYTHIA. For all other technical details we refer to \cite{Buss:2011mx,gibuu}. No parameters in this model are tuned to neutrino data, with one exception: the $2p-2h$ hadron tensor, assumed to be purely transverse, has its strength fitted to the semi-inclusive MiniBooNE double-differential no-pion data \cite{Lalakulich:2012ac}. The oscillation parameters were chosen to be the same as those used by Bishai et al. \cite{Bishai:2012ss}.

\begin{figure}
\includegraphics[angle=-90,width=0.5\textwidth]{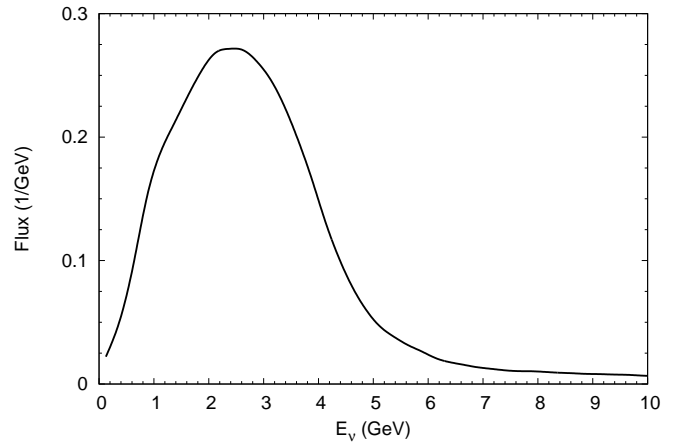}
\caption{Flux distribution for LBNE, normalized to 1 in the full energy regime $0 \le E \le 30$ GeV.  The distribution is taken from \cite{LBNEDES:2012}.} \label{fig:LBNEflux}
\end{figure}

For our analysis we assume GiBUU to be "nature" and generate full events for true energies which we distribute according to the LBNE flux \cite{LBNEDES:2012} shown in Fig.\ \ref{fig:LBNEflux}; the target is $^{40}$Ar.
Using only the muon kinematics we then reconstruct the neutrino energy using the expression appropriate for true QE scattering from a neutron at rest
\begin{equation}
E_{\nu} = \frac{2(M_n - E_B)E' - (E_B^2 - 2 M_nE_B + m_\mu^2 + \Delta M^2)}{2(M_n - E_B - E' + |\vec{k}'| \cos \theta')} ~.
\end{equation}
Here $\vec{k}'$ is the outgoing lepton's momentum, $E'$ its energy and $\theta'$ its angle. The quantity $E_B$ denotes an average binding energy of the neutron inside the nucleus; it is taken to be $E_B = 0.03$ GeV. Furthermore, $\Delta M^2 = M_n^2 - M_p^2$.

It is customary to remove from this event sample all events with pions in the final state. This selection eliminates a large part of the resonance excitation and DIS processes, but  events with initial pion or $\Delta$ resonance production and subsequent reabsorption are still contained in the sample. Cherenkov detector experiments use this method to identify QE-like events. Since the LBNE plans to use LAr detectors, which allow tracking of charged particles, we also study a second alternative that further restricts the event sample. Our studies of various observables in \cite{Lalakulich:2012ac} for the energy regime between about 0.5 and 2 GeV and a {\it C} target we had shown that events with 0 pions, exactly 1 proton and X (unobserved) neutrons were dominated by QE \footnote{Events with 0 pions, exactly 1 proton and no other baryons were found to be nearly clean QE events, but present difficulties because the neutrons often are not detected.}. We, therefore, here also employ this restriction in addition to explore its influence on the energy reconstruction also at the higher energies of the LBNE. For the theoretical analysis inclusive cross sections are not sufficient, but full events first have to be generated.

\section{Results}

\begin{figure}
\includegraphics[angle=-90,width=0.85\textwidth]{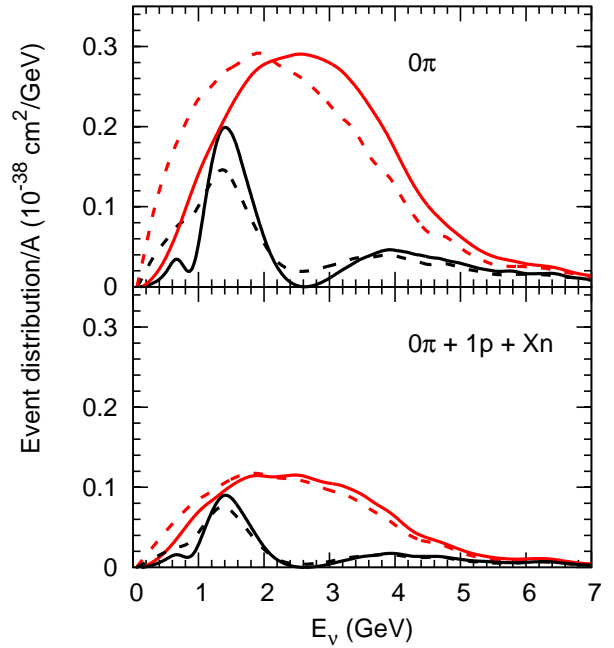}
\caption{(Color online) Event distribution (normalized flux times cross section) per nucleon for LBNE vs.\ true (solid curve) and reconstructed (dashed curve) energy. The upper two (red) curves give the distribution without oscillation, the lower two (black) curves give the distribution with oscillation in the muon disappearance channel. In the upper part of the figure  the events have no pions in the final state, in the lower part the events have 0 pions, exactly 1 proton and $X$ neutrons in the final state.} \label{fig:events}
\end{figure}

In the upper part of Fig.\ \ref{fig:events} we show first the distribution for 0-pion events both at a near detector, without oscillations, and at the far detector, with oscillations, in the muon disappearance channel.
There is a dramatic shift in energy visible in the unoscillated (upper) curves; the event distribution plotted vs.\ reconstructed energy is tilted by about 0.5 GeV towards lower energies, compared to the distribution as a function of true energy.  At the peak of the distribution about 50\%  of the total comes from true QE events and about 20\% from $\Delta$ excitation. The remainder comes to about equal parts from 2p-2h excitations and from DIS events. The event rates after oscillation are given by the lower two curves. Even the reconstructed event distribution (dashed) clearly shows the oscillation signal, but again it is distorted. The main effect of the energy reconstruction is a filling in and flattening of the minimum around 2.7 GeV, together with a significant lowering of the second maximum at around 1.4 GeV. The dramatic shift in the unoscillated distributions is replaced by a considerable broadening of the oscillation maxima in the distribution plotted vs.\ reconstructed energy (dashed curves).  At around 1.4 GeV the two solid curves as a function of true energy coincide whereas those as a function of reconstructed energy (dashed) are quite different. This difference is due to the fact that the measured event distribution depends on the reconstructed energy which, at a fixed value, corresponds to a superposition of many, mainly larger, true energies (cf.\ Fig.\ 6 in \cite{Lalakulich:2012hs}).

The lower part of Fig.\ \ref{fig:events} shows the same quantities, but now obtained for a more restricted event sample of 0 pions, exactly 1 proton and $X$ neutrons. One sees that now the solid and dashed curves, i.e.\ the true and reconstructed results, agree much better with each other. The downward shift in the reconstruction is still visible, but significantly decreased to about 100 MeV. This comes at the expense of the number of events that is now about a factor of 3 lower than that for the 0-pion events. Closer inspection of the events shows that now about 80\% originate in true QE, with contributions from $\Delta$ excitation, $2p-2h$ and DIS accounting to about equal parts for the rest (at the peak of the distribution). The true QE events have thus significantly been enhanced.

By varying the $2p-2h$ contribution which is so far only restricted by the MiniBooNE data we have verified that these results are quite robust and do not depend on the specifics of our treatment of the $2p-2h$ excitation. This can be understood since a charged current reaction favors initial pp production and fsi tend to increase the number of final state nucleons.

\begin{figure}
\includegraphics[angle=-90,width=0.5\textwidth]{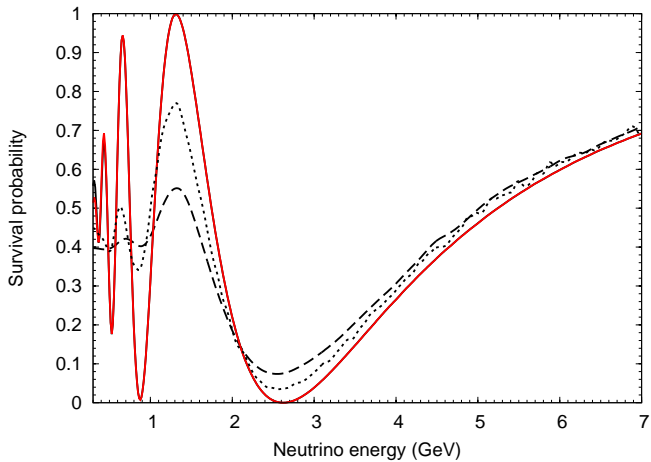}
\caption{Survival probability for $\mu$-neutrinos both for true (solid) and reconstructed (dashed) energies for 0-pion events. The dotted curve gives the probability for events with 0 pions, 1 proton and $X$ neutrons.}
\label{fig:oscillfact}
\end{figure}

Experimentally, the oscillation probability will be obtained by dividing the oscillated (far-detector) by the unoscillated (near-detector) flux. The result is shown in Fig.\ \ref{fig:oscillfact}, again both as a function of true and of reconstructed energy.
The main effect caused by the inherent errors in the reconstruction is now a significant change in the absolute values both at the first minimum and the first maximum whereas the locations of these two points are less affected. This change will manifest itself in a significant error in the mixing angle whereas the $\Delta m^2$ is less modified. This same effect can also be seen in the results of Coloma et al. for T2K \cite{Coloma:2013rqa} which show that the mixing angle is significantly more affected by uncertainties in the nuclear model than the squared neutrino mass difference.

The dotted curve in Fig.\ \ref{fig:oscillfact} gives the same reconstructed survival probability but now obtained from the event sample with 0 pions, 1$p$ and $Xn$. While the position of maxima and minima in comparison to the true survival probability (lower solid curve) is only insignificantly affected the values at these points are now much better reproduced than in the results for the 0 pion sample (dashed curve).

\begin{figure}
\includegraphics[angle=-90,width=0.5\textwidth]{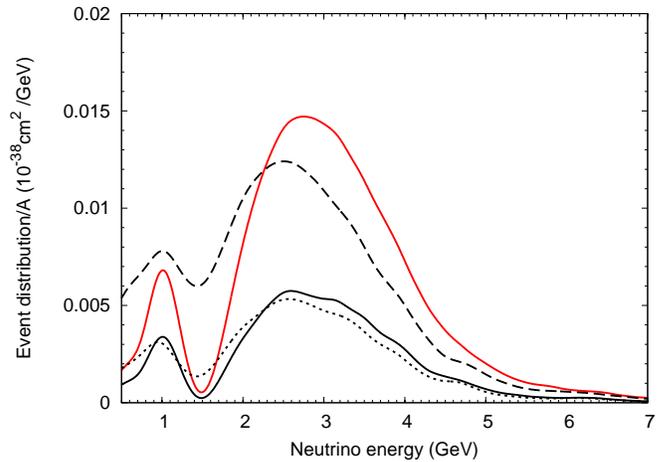}
\caption{Event distribution per nucleon for electron appearance both for true (solid) and reconstructed (dashed) energies. The $CP$-violating phase has been set to 0. The upper two curves are based on 0-pion events, the lower two curves on events with 0 pions, 1 proton and $X$ neutrons.} \label{fig:e-app}
\end{figure}

We have also looked at the sensitivity of the electron appearance signal to the energy reconstruction \footnote{For this study we have neglected the electron-muon mass difference. The results are therefore unreliable for small neutrino energies close to the muon mass. This shortcut allowed us to use for this analysis the 0-pion events generated with an incoming muon neutrino beam, thus resulting in savings of computer time. }. The comparison is shown in Fig.\ \ref{fig:e-app} for a $CP$-violating angle $\delta_{CP} = 0$.
For the event sample with the 0-pion restriction the main effects are again a shift of the reconstructed event distribution by about 500 MeV towards lower energies and a significant filling in of the minimum around 1.4 GeV. The latter reflects the significant admixture of larger true energies at a fixed reconstructed energy. Again, for the more restrictive event sample with 0 pions, $1p$ and $Xn$ (shown in the lower part of Fig.\ \ref{fig:e-app}) the true and reconstructed curves are quite close to each other (lower solid and dotted curve, resp.). In particular the oscillatory structure is quite well reproduced and the remaining shift amounts to less than about 100 MeV. As for the disappearance the loss of events amounts to about a factor of 3.

\begin{figure}
\includegraphics[angle=-90,width=0.85\textwidth]{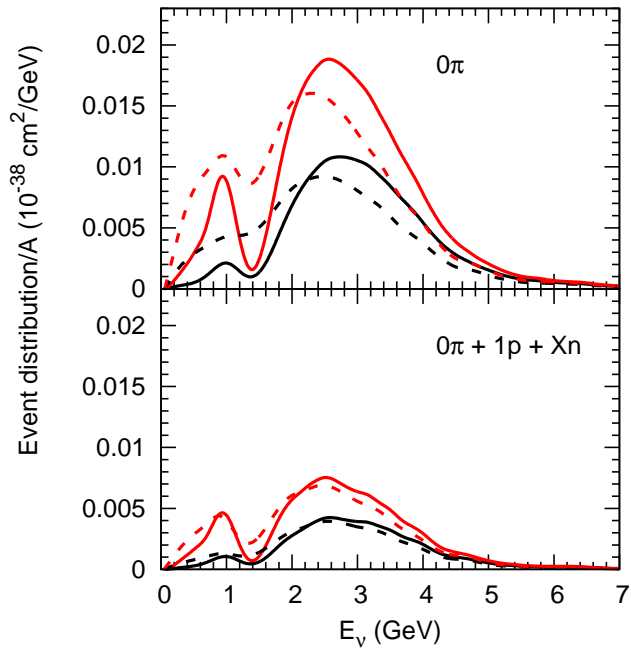}
\caption{(Color online) Event distributions per nucleon for electron appearance for $\delta_{CP}=+ \pi/2$ and $\delta_{CP}=-\pi/2$ (upper red and lower black curves, resp.), both for true (solid) and reconstructed (dashed) energies. The upper part gives the results for 0 pion events, the lower part gives the results for events with 0 pions, 1 proton and $X$ neutrons.} \label{fig:deltaCP}
\end{figure}

Particularly interesting is the sensitivity of this signal to the presence of a nonvanishing $CP$-violating phase $\delta_{CP}$ and its dependence on the energy reconstruction. This is shown in Fig.\ \ref{fig:deltaCP} for the two extreme cases $\delta_{CP} = \pm \pi/2$.
For $\delta_{CP} = -\pi/2$ the minimum at around 1.5 GeV has now nearly completely disappeared in the distribution vs.\ reconstructed energy for the 0-pion events in the upper part of Fig.\ \ref{fig:deltaCP}. The differences between the event distributions for true and reconstructed energy are particularly large to the left of the main peak. However, the further restriction of the event sample to 0 pions, 1 proton and $X$ neutrons changes this picture dramatically (see lower part of Fig.\ \ref{fig:deltaCP}). Now again the true and reconstructed curves have a very similar structure with a shift of only about 100 MeV.

\section{Summary and Conclusions}
Previous studies of the physics potential of the LBNE have illustrated that an energy resolution of about 100 MeV is necessary to distinguish between different physics properties \cite{Adams:2013qkq}. The present investigation has shown that a QE-based energy reconstruction, using a 0-pion event sample, is subject to errors of up to 500 MeV in the neutrino event rates as a function of energy. Correspondingly, the oscillation signal for muon disappearance exhibits large uncertainties in the region of the first minimum and second maximum. The QE-based energy reconstruction method using a 0-pion event sample can thus not reach the necessary accuracy. In principle, reconstructed energies could be transformed back to a distribution of true energies by means of migration matrices calculated with a neutrino generator. Since generators have their own uncertainties this migration from reconstructed to true energy is the more reliable the closer the  reconstructed energy is already to the true energy. A difference of 500 MeV is too large for this and any migration would introduce additional generator dependence into the data.

A major improvement takes place when the event sample is further restricted by the requirement of 0 pions, exactly 1 proton, and $X$ (unobserved) neutrons. In this case the shifts between reconstructed and true event rates drop to about 100 MeV and thus become close to the required energy resolution. This result depends crucially on the fact that events with 0 pions and only 1 outgoing proton are primarily due to an original QE event; that the latter is true we had already shown for the lower-energy MiniBooNE flux and a lighter target ({\it C}) \cite{Lalakulich:2012ac}. We expect it to be generally true because events with 0 pions and only 1 proton are produced predominantly in QE events. While one loses only about a factor of 3 in the number of events, the accuracy of the reconstruction gained by this restricted event sample is impressive. We thus conclude, that even for a higher-energy experiment such as LBNE the energy reconstruction can be based on the QE method, if the event sample is properly chosen.

This choice is by experimental means only, no generator is needed for it. Only the migration back from reconstructed to true energy requires a generator, but now the difference between the true and the reconstructed distributions is much smaller and the generator dependence is minimized. We also note that this applies also to the ongoing MINER$\nu$A experiment which works with a similar flux, as well as any experiment that can do tracking of protons.

It will be interesting to perform a similar study for the calorimetric method; here only some early studies exist \cite{Harris:2003si,Kordosky:2007tu,Dytman:2008zz} which indicated an inaccuracy of about the same order as the one found here for the restricted event sample. With better event generators a more accurate description of the reconstruction of the energy in the calorimetric method should be possible. The invisible part of the energy, e.g., in the form of stuck pions, can nowadays rather reliably be modeled in generators that have been checked with pion production data.  For the visible hadronic shower energy an experimental acceptance filter for the outgoing particles would be most useful for further simulations. It could then be imposed on a full event sample produced by GiBUU, available from \cite{gibuu}, to obtain an expected experimental signal that could directly be compared with the calculated true event distributions. It remains to be seen if this method can yield better accuracies than the QE-based method discussed here.

\begin{acknowledgments}
We are grateful for the help of the whole GiBUU team in developing both the physics and the code used here.
One of the authors (U.M.) also gratefully acknowledges stimulating discussions with P.\ Huber and C.\ Mariani. We also thank P.\ Huber for providing us with the LBNE flux.

This work has been supported by BMBF.
\end{acknowledgments}

\bibliography{nuclear}

\end{document}